# Temperature dependent network stability in simple alcohols and pure water: the evolution of Laplace spectra

*I. Bakó[1], I. Pethes[2], Sz. Pothoczki[2], L. Pusztai[2,3]*

[1] Research Centre for Natural Sciences, Hungarian Academy of Sciences, H-1117 Budapest, Magyar tudósok körútja 2., Hungary
[2] Wigner Research Centre for Physics, Hungarian Academy of Sciences, H-1121 Budapest, Konkoly Thege M. út 29-33., Hungary
[3] International Research Organization for Advanced Science and Technology (IROAST), Kumamoto University, 2-39-1 Kurokami, Chuo-ku, Kumamoto 860-8555, Japan

## Abstract

A number of computer-generated models of water, methanol and ethanol are considered at room temperature and ambient pressure, and also as a function of temperature (for water and ethanol), and the potential model (for water only). The Laplace matrices are determined, and various characteristics of this, such as eigenvalues and eigenvectors, and the corresponding Laplace spectra are calculated. It is revealed how the width of the spectral gap in the Laplace matrix of H-bonded networks may be applied for characterising the stability of the network. A novel method for detecting the presence percolated network in these systems is also introduced.

## 1. Introduction

Network theory is a useful tool for analyzing complex systems. In a complex system, there are large numbers of components that interact with each other. Networked structures appear in a wide variety of different contexts, such as technological (large communication systems, Internet, telephone networks), transportation-related, social phenomena and biological systems [1-10]. In these studies of networks where large systems of interactions are mapped into graphs, vertices and edges are usually recognized as the primarily building blocks. Given a graph we can associate several matrices which record information about vertices and how they are interconnected.

Over the past several years considerable attention has been focused on studying complex networks. A large number of concepts have been introduced for classifying network structures, for example degree distribution, path length, clustering, percolation, the small world property, spectral density of graph, ring structures, etc. During the last years it has become clear that the Laplacian eigenvalues and eigenvectors play an important role in revealing the multiple aspects of the characteristics of network structure and dynamics, like spanning trees, resistance distances and community structures [11-22].

There are several example of application of network science concepts in chemistry and physical chemistry, too [23-33]. Hydrogen-bond (HB) connectivity and strength and directionality influence the anomalous properties of water and other H-bonded liquids and liquid mixtures. It is widely accepted that water prefers to form a complicated three-dimensional percolated network [34-36]. On the other hand, bulk methanol and ethanol form (mostly) chain-like clusters [38,39]. Recently we have applied well known graph theoretical concepts for revealing basic units of the network structure in liquid water at room temperature, in water-methanol mixtures at different temperatures [40,41], and in formamide [42]. We found that methanol molecules prefer to form non-cyclic entities, whereas water molecules favour to build rings and the number of cyclic entities in the mixtures is increasing as the temperature is decreasing.

Several authors have already used the properties of the spectral density of adjacency and/or Laplace matrices calculated for H-bonded networks, in order to characterise more deeply the H-bonded structure in pure liquids and liquid mixtures [24-30,33] . In the current study, we investigate spectral properties of the hydrogen bond network in liquid water, as well as in simple alcohols methanol and ethanol at different temperatures. At present, we focus mainly on producing some metrics that could provide estimates as to how far our systems are from a percolation transition. Establishing such metrics would be extremely helpful for identifying percolating/non-percolating systems without lengthy 'self-avoiding walk' type computations on large particle configurations.

## 2. Network theory background and calculation details

### 2.1 Network Spectra

The structure of complex networks mentioned above can be completely described by the associated adjacency (A), combinatorial Laplace (L) matrices. The adjacency matrix is defined by $A_{ij}=1$ if an edge (bond) exists between the nodes numbered i and j and $A_{ij}=0$ otherwise. The Laplace matrix can be defined as follows:

$$L_{ij} = k_i \delta_{ij} - A_{ij} \qquad (1)$$

where $k_i$ is the number of (hydrogen) bonded neighbors of particle 'i'.

The spectrum of a graph is the set of eigenvalues of its corresponding matrices (A, L). The properties of eigenvalues and eigenvectors of these matrices are related to the inherent properties of networks. The eigenvalues of adjacency matrices were investigated in more detail in the past than the eigenvalues of the Laplacian matrices. However, it has already been shown that the solution of the diffusion and flow problems on networks (spreading diseases, random walks, navigated random walks [1-9]) is related to the Laplacian spectra.

There are many interesting results known about Laplacian matrices. *L* is positive semidefinite and has nonnegative eigenvalues. Furthermore, 0 is always an eigenvalue of *L* and the multiplicity of the eigenvalue 0 is equal to the number of connected components of the graph. Many authors have studied the relationship between the eigenvector corresponding to the second smallest eigenvalue ($\lambda_2$) and the graph structure; well documented reviews can be found in the literature [11-15]. $\lambda_2$ can also be called as the 'spectral gap', since its value provides directly the difference (the 'gap') between the smallest (always zero, if the graph is connected) and the second smallest eigenvalues (Fiedler eigenvalue). One of the frequently used theorems connected to the second smallest positive eigenvalue of a Laplacian is known as the Cheeger inequality [22]:

$$\frac{\lambda_2}{2} \leq h(G) \leq \sqrt{2\lambda_2} \qquad (2)$$

where h(G) is the Cheeger constant of a graph G. h(G) is defined by the following equation

$$h(G) = \min \frac{E(S,S_c)}{\min(vol(S),vol(S_c))} \qquad (3)$$

Here S, $S_c$ are two non-empty subsets of G, vol(S) is the sum of the edges, and $E(S,S_c)$ is the sum of the edges within S and $S_c$. This inequality is related to the minimum number of edges such that, when removed, cause the graph to become disconnected ('non-percolated' in the world of H-bonded networks) [11-14]. h(G) (or a similar derived quantity like $\lambda_2*n_{hb}$) can therefore serve as possible metrics to measure the 'distance' from the percolation transition.

Methods for the identification of the percolation transition in pure liquid and liquid mixtures have already been described in several works. One of the most widely used forms relies upon the cluster distribution ($n_s$) which deviates from a power law form and decreases much faster with s than $n_s \sim s^{-2.19}$ [43], as it is expected for random percolation on a 3D cubic lattice. The so-called Fiedler eigenvalue has been extensively studied in diverse application areas, including biology, neuron science, spectral clustering technique, etc... [18-22, 43-45].

It has already been shown that the presence of near-zero eigenvalues generally indicates the existence of strong communities, or nearly disconnected components [10,14]. Another prominent feature for the eigenvalue problem of the L matrix is revealed by the structure and localization of components of the eigenvectors. We can characterize these eigenvectors ($\lambda_j$) by using the inverse participation ratio (I), as defined by the following equation

$$I_j = \sum_i V_{ji}^4 \qquad (4)$$

where $V_{ji}$ is the i-th element of the j-th eigenvector. It has already been also shown that 'I' ranges from the minimum value of 1/N, corresponding to the eigenvector distributed equally on all nodes, to a maximum of 1 for a vector with only one nonzero component.

Similarly, we can characterize the average localization of the i-th node (NPR , nodes participation ratio):

$$D_i = \sum_j V_{ji}^4 \qquad (5)$$

where the summation is over the i-th components of the eigenvectors. Note that $D_i$ is 1 in the case particle 'i' is a monomer.

## 2.2 Computer-generated structural models

Most structures considered here have been prepared in-house, for the purpose of the present investigations. All of the molecular dynamics simulations, using rigid non-polarisable water models (SPC [44], SPC/E [43], TIP3P [45], TIP4P [46]), have been performed by the GROMACS simulation package [50] (version 5.1.1). The leap-frog algorithm has been used for integrating Newton's equations of motion, with a time step dt=1 fs, at the experimental density of water at 298 K, with N=2048 water molecules. We performed additional simulation using the polarizable Amoeba [47] water model, with the same simulation parameters, using the Tinker (version 8.4) code [49]. The simulation length was about 5 ns, after a 1 ns equilibration period. We calculated the Laplace spectra of the H-bonded network from 100 configurations, taken equidistantly over the last 5 ns. The Laplace spectra of the H-bonded network from an AIMD simulation [48] (N=1024, T=315 K, BLYP-D3) was calculated for only a single frame, which was the last step of a 10 ps simulation.

## 3. Results and discussion
### 3.1 Laplace spectra

Fig. 1 depicts the Laplace spectra of the investigated graphs for pure methanol and ethanol. For ethanol we also show how the spectra change as a function of temperature. The spectra of methanol and ethanol has several well-defined peaks ($\lambda$=0.5, 1, 1.5, 2, 3…). Since these are rather characteristic features, in the future it may be worthwhile investigating their significance and possible meaning

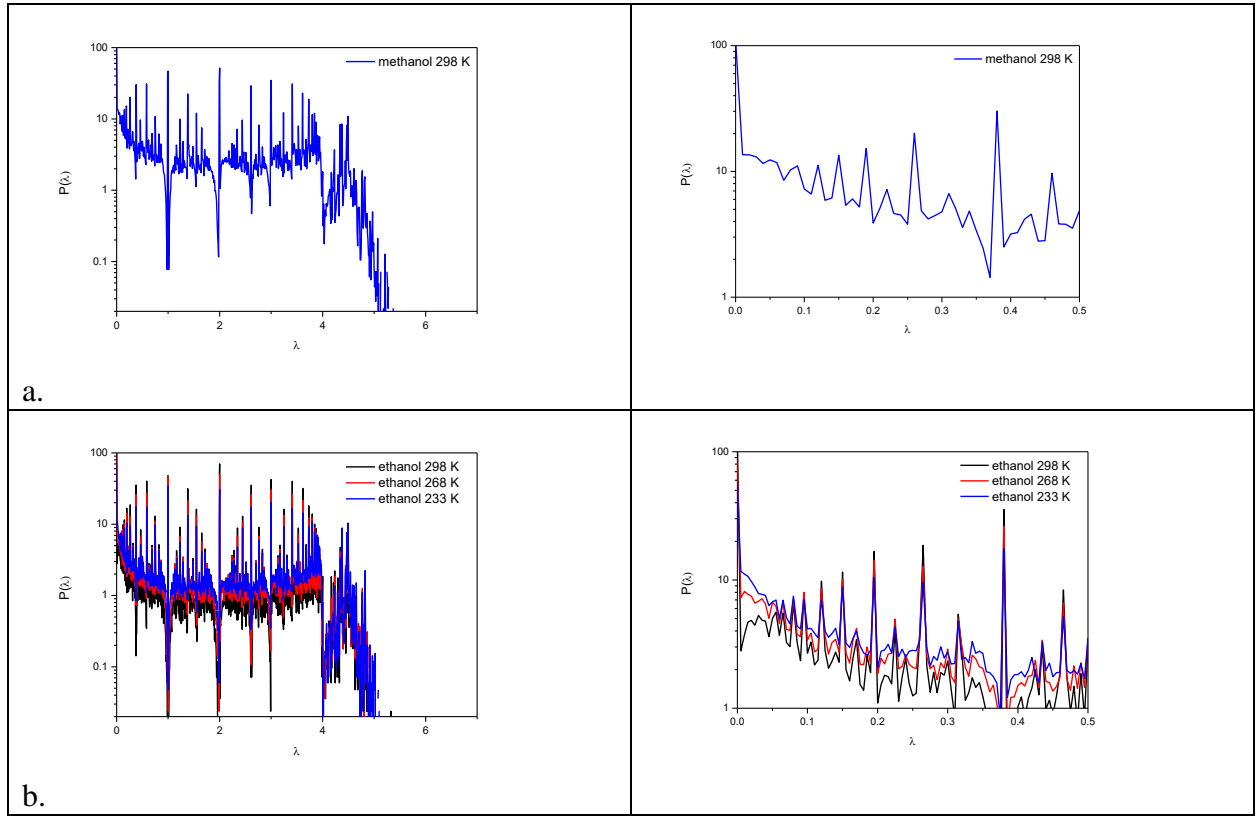

Fig 1. Laplace spectra of H-bonded networks in liquid methanol (a) and ethanol (b) at different temperatures. On the right side we showed the Laplacian spectra in a smaller range near to 0.0.

We can conclude from the investigation of the behavior of the spectra at low λ value that these spectra do not indicate any recognizable gap (i.e. a value of the second eigenvalue that is clearly distinct from zero). It follows from these results that liquid methanol and ethanol do not form percolated networks (not even at low temperature); this is in good agreement with previous results [38,39]. In order to prove this statement more precisely, we investigated the distribution of the inverse participation ratio (IPR) for the 0 eigenvalue. As mentioned earlier, the number of 0 eigenvalues equals to the number of connected components in the network. If the IPR of one of the eigenvalues is small (in the order of 1/N) it means that the network consists of N nodes (molecules) We can reach this conclusion also by investigating the right side (large cluster size) of the cluster size distributions. Fig. 2 shows the IPR for liquid ethanol, together with their corresponding cluster size distributions.

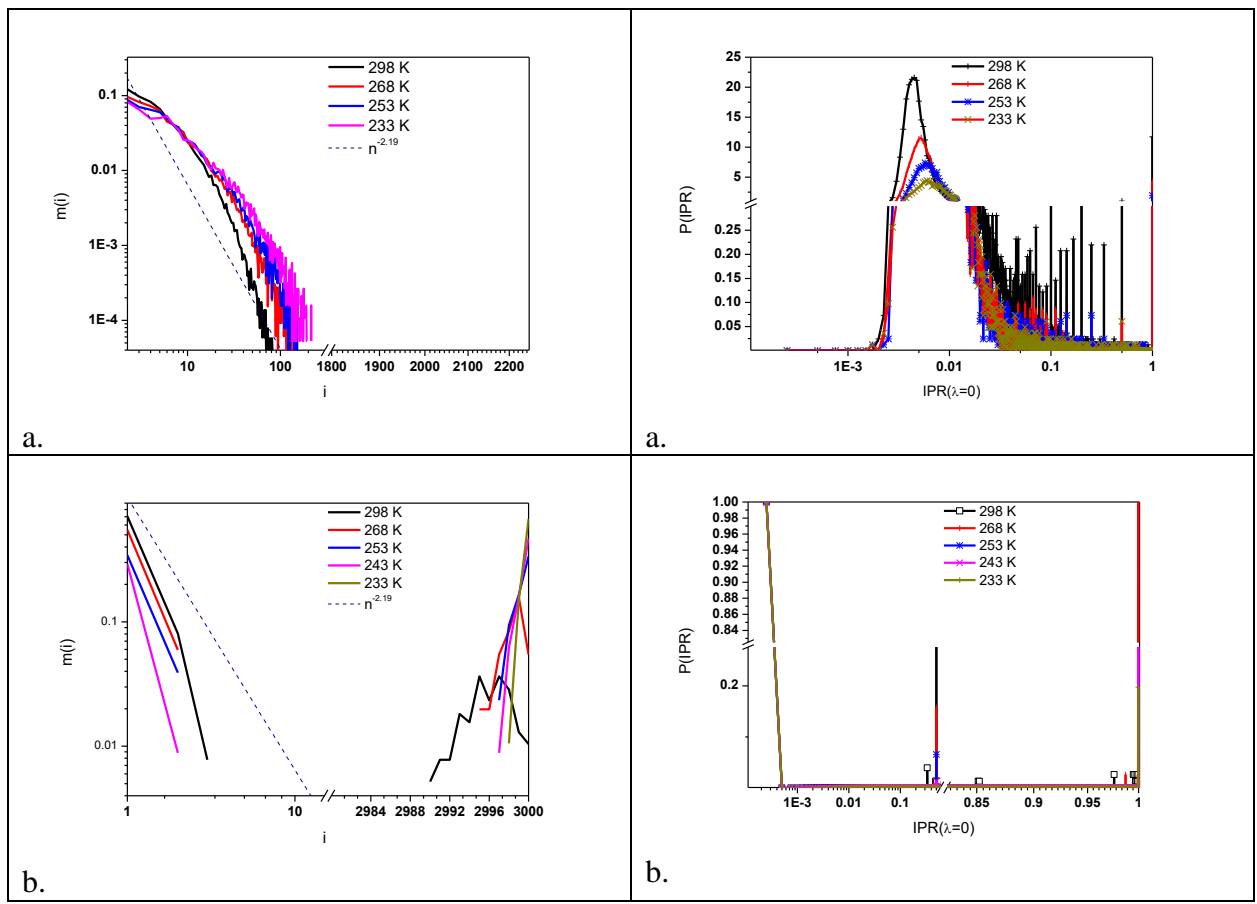

Fig. 2. Cluster size distributions and Inverse Participation ratios (IPR) of the '0' eigenvalue for ethanol(a) and (SPC/E)(b) water at different temperatures.

The IPR distributions for liquid ethanol (Fig 2. b) have well defined maxima around 0.007-0.01, which corresponds approximately to association with 10-15 ethanol molecules bonded together via H-bonds. From the investigation of the cluster size distribution for liquid ethanol (Fig. 2, a) we can also conclude that there is not any percolated network in the system (no bump at large cluster size) and that the average cluster size is approximately 10-15.

The IPR distributions for liquid water have well defined maxima around $5*10^{-4}$ that correspond to delocalization over the entire network. This agrees well with the conclusion from the cluster size distribution, namely that liquid water does have a percolated network.

Fig 3. Laplace spectra of the H-bonded network in liquid water (SPC/E model at different temperatures). On the right side the same Laplacian spectra are shown for a narrow range of small eigenvalues.

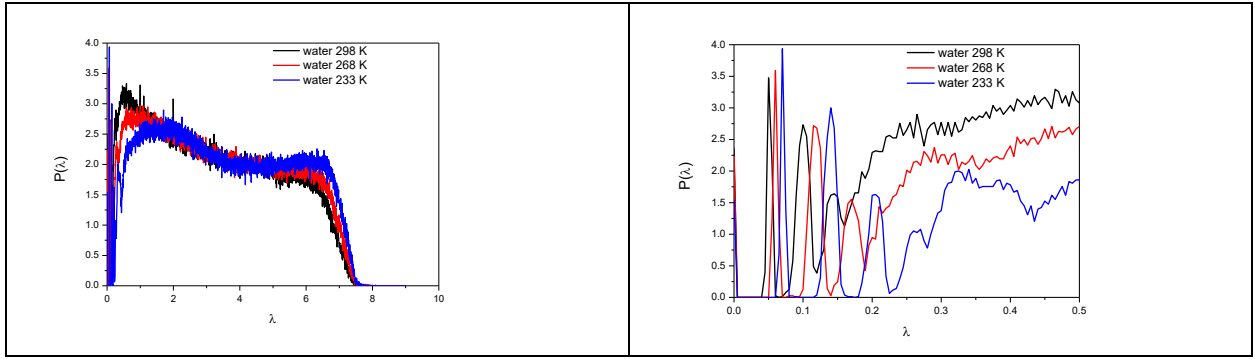

We present the Laplace spectra for water (SPC/E model) at different temperatures in Fig. 3. We can clearly detect a well-defined gap at some low eigenvalue that becomes smaller as temperature increases. The presence of near-zero eigenvalues indicates the existence of strong communities or nearly disconnected components. We can use the position of the first peak (according to the Cheeger inequalities, see eq.2 ) of these graphs as metrics for describing the "distance" (number of bonds that need to be broken) from the percolation transition. Further maxima appear in this figure that become better defined at lower temperatures. Investigation of the IPR and the cluster size distribution (shown in Fig 2.) also revealed that in these systems we have percolated networks. It should be noted that similar Laplace spectra can be obtained from several types of classical and ab initio models of liquid water, as represented in Fig 4. (most data are shown for room temperature). Characteristic properties of different water models are presented in Table 1. We can conclude that the widest spectral gap is found for the Amoeba, while the smallest one is for the TIP3P model. These values are correlated well with their average H-bond number.

Fig 4. Laplace spectra of H-bonded networks in liquid water using different water model. (a) SPC, SPC/E, TIP4P, TIP3P, and Amoeba at 298 K; (b) AIMD simulation at 315 K.

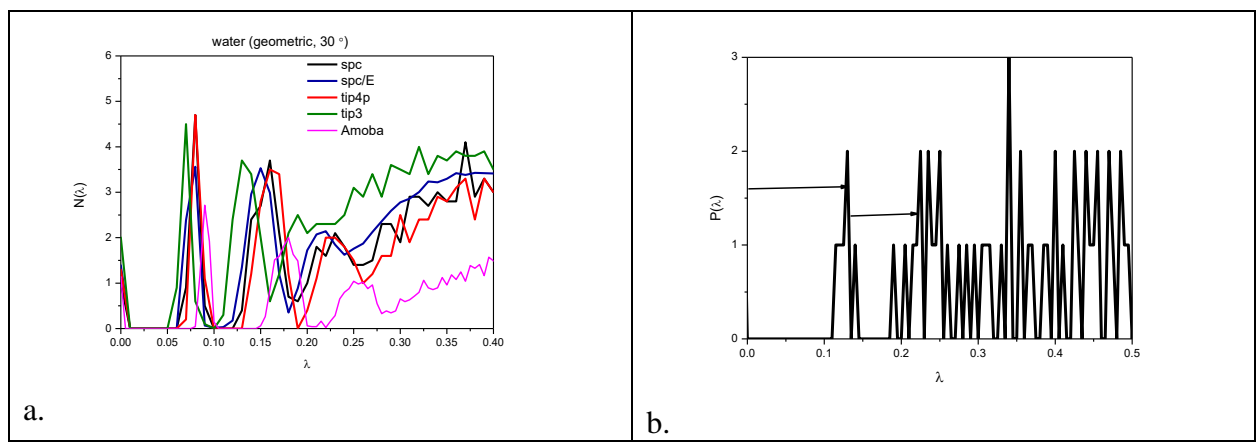

Table 1. Characteristics of the H-bonding network for various water models.

|  | T(K) | $n_{hb}$ | $\lambda_2$ | $\lambda_2 \cdot n_{hb}$ |
|---|---|---|---|---|
| SPC/E(43) | 298 | 3.174 | 0.050 | 0.159 |
|  | 268 | 3.340 | 0.060 | 0.200 |
|  | 253 | 3.426 | 0.062 | 0.212 |
|  | 243 | 3.489 | 0.065 | 0.227 |
|  | 233 | 3.551 | 0.071 | 0.252 |
| SPC(44) | 298 | 2.870 | 0.040 | 0.115 |
| TIP4P/E(46) | 298 | 3.190 | 0.064 | 0.204 |
| TIP3P(45) | 298 | 2.730 | 0.040 | 0.109 |
| AIMD(48)* | 315 | 3.450 | 0.090 | 0.311 |
| Amoeba (47)* | 298 | 3.470 | 0.090 | 0.312 |
| SPC* | 298 | 3.350 | 0.050 | 0.168 |
| SPC/E* | 298 | 3.410 | 0.060 | 0.205 |
| TIP4P/E* | 298 | 3.570 | 0.072 | 0.257 |
| TIP3P* | 298 | 3.260 | 0.040 | 0.130 |

∗ geometric definition $r_{OH} < 2.5$ Å and HOH $\angle < 30°$; in other cases we used an energetic definition for detecting HB-s in the system as $r_{OH} < 2.5$ Å and $E_{hb} < -3.0$ kcal/mol

## 3.2 Localization of the nodes

The distribution of localization functions of the nodes, P(NPR), are presented in Fig. 5. (a. water, b:ethanol)

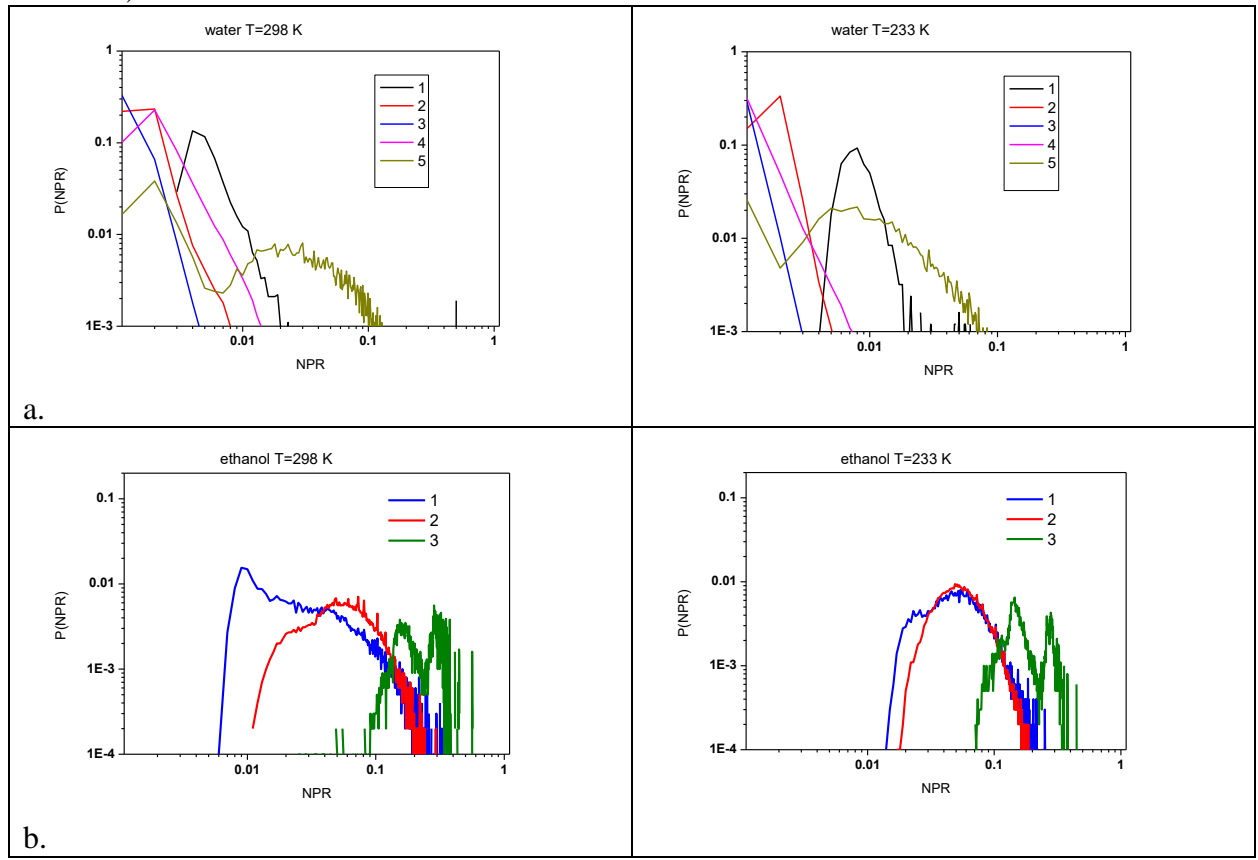

Fig. 5. Localizations of nodes (P(NPR)) as a function of H-bond number for water and ethanol at two different temperatures

It is clear from these figures that the localization of water molecules with 1 and 5 hydrogen bonded neighbors is significantly different from the other structures. The difference becomes better defined for water molecules with 1 H-bonded neighbor at low temperature. This function for water molecules with 2, 3 and 4 hydrogen bonded neighbors has short tail like distributions. We have already found this type of distribution for small world graph 'h' with p=0.4 rewiring probability [33]. From the tree like structure we see for ethanol the existence of different localizations of 1, 2, or 3 hydrogen bonded ethanol molecules can be confirmed. Mainly all of the NPR values for ethanol molecules are larger than the same values for water. We can detect a significant change as a function of temperature for ethanol molecules with 1 H-bond.

All the above results point to the consequence that the H-bonded network in water—in good agreement with the earlier studies – is significantly more connected and complex than in liquid ethanol.

## 4. Conclusions

In this work we showed how the size of the spectral gap in the Laplace matrix of H-bonded networks may be applied for characterising the stability of the network. This quantity does not depend on the applied water model (at least among the ones considered here). We introduced a new method for detecting the presence percolated network in these systems, too, via the investigation of the properties of the distribution of inverse participation ratio for eigenvalue '0'. We showed that the localisation of water molecules that belong to the network is significantly different from the ones that are not part of the percolated cluster.

**Acknowledgments**

The authors are grateful to the National Research, Development and Innovation Office (NRDIO (NKFIH), Hungary) for financial support via grants Nos. SNN 116198 and 124885.